\documentclass[twocolumn,showpacs,preprintnumbers,amsmath,amssymb]{revtex4}
\usepackage{graphicx}
\usepackage{dcolumn}
\usepackage{bm}

\begin{document}

\title{Pulse trapping inside one-dimensional photonic crystal with relaxing cubic nonlinearity}

\author{Denis V. Novitsky}
\email{dvnovitsky@tut.by}
\affiliation{%
B.I. Stepanov Institute of Physics, National Academy of Sciences of
Belarus, \\ Nezavisimosti~Avenue~68, 220072 Minsk, Belarus.
}%

\date{\today}

\begin{abstract}
We theoretically study the effect of pulse trapping inside
one-dimensional photonic crystal with relaxing cubic nonlinearity.
We analyze dependence of light localization on pulse intensity and
explain its physical mechanism as connected with formation of
dynamical nonlinear cavity inside the structure. We search for the
range of optimal values of parameters (relaxation time and pulse
duration) and show that pulse trapping can be observed only for
positive nonlinearity coefficients. We suppose that this effect can
be useful for realization of optical memory and limiting.
\end{abstract}


\maketitle

\section{Introduction}

Photonic-band-gap structures are actively studied as promising
elements for different devices of nonlinear and quantum optics
\cite{Yabl, Bert, Joan}. Study of nonlinear photonic crystals is
connected with the possibility of dynamical adjustment of the
parameters of the system. Photonic crystals (including two- and
three-dimensional) allow to control dispersion and diffraction
properties of light \cite{Stal, Chutinan}, obtain pulse reshaping
\cite{Vujic}, pulsed high-harmonic generation \cite{Scalora}, etc.
The processes of light self-action in such systems result in
localization effects such as gap solitons formation, discrete mode
existence or light energy localization (see, for example,
\cite{Ming00, Ming01, Ming02, Sukh}).

The simplest example of one-dimensional photonic crystal can be
represented as a periodic set of alternating dielectric layers with
large depth of refractive index modulation. One of the most
prominent effects of nonlinear optics of such structures is strong
pulse compression in photonic crystals with non-resonant cubic
nonlinearity \cite{Eggl, Zhelt}. The effect of light pulse
localization, or trapping, in one-dimensional photonic crystal with
a defect was studied in Ref. \cite{Good} and, in more general form,
in Refs. \cite{Mak1, Mak2}.

In this paper we study pulse propagation in photonic crystal with
relaxing cubic nonlinearity. It is obvious that if we reduce
incident pulse duration, the inertial properties of medium
nonlinearity should be taken into account. Indeed, the lowest values
of relaxation time connected with the electronic Kerr mechanism (a
few femtoseconds \cite{Akhm}) appear to be comparable with pulse
durations obtained at the modern set-ups. As it was shown in Ref.
\cite{Vlasov}, relaxation of nonlinearity results in vanishing of
the effect of femtosecond pulse compression which can be obtained in
relaxation-free case.

In our research we use numerical simulations of the Maxwell wave
equation taking into account the process of nonlinearity relaxation.
The method used allows to obtain numerical solutions of the problem
without any assumptions about medium parameters modulation or rate
of variation of field envelope. We show that, in a certain region of
pulse amplitudes, relaxation times and pulse durations, the pulse
can be trapped inside a nonlinear cavity dynamically formed by the
light. Appearance of the cavity is connected with local change of
reflective properties of the photonic structure. In other words,
dynamical shift of band spectrum occurs. Trapping pulse inside the
photonic crystal is a prospective effect for such possible
applications as optical limiters (which transmit light only with
proper intensity), optical buffers or memory (which allows to store
light for time, large in comparison with characteristic transmission
time). Large advantage of the scheme considered is absence of
necessity to introduce nonuniformity or any imperfections in the
structure of nonlinear photonic crystal.

The article is divided into several sections. In Section \ref{SecEq}
the main equations are given and the approach for numerical solving
the wave equation is considered. Section \ref{SecTrap} is devoted to
some phenomenological aspects of pulse trapping effect. In Section
\ref{SecMech} the physical mechanism of pulse trapping in photonic
crystal with relaxing cubic nonlinearity is discussed. Finally, in
Section \ref{SecDep} we consider some conditions for pulse trapping
observation connected with the proper choice of pulse duration and
relaxation time.

\section{\label{SecEq}Main equations and numerical method}

In this paper we consider ultrashort pulse interaction with
one-dimensional photonic crystal made of substance with relaxing
cubic (Kerr) nonlinearity. Light propagation along $z$-axis is
governed by the Maxwell wave equation
\begin{eqnarray}
\frac{\partial^2 E}{\partial z^2}&-&\frac{1}{c^2} \frac{\partial^2
(n^2 E)}{\partial t^2} = 0, \label{Max}
\end{eqnarray}
where $E$ is electric field strength, $n$ is medium refractive index
that depends on light intensity $I=|E|^2$ as
\begin{eqnarray}
n=n_0(z)+\delta n (I, t). \label{refr}
\end{eqnarray}
Here $n_0(z)$ is a linear part of refractive index. Time dependence
of nonlinear term $\delta n$ is responsible for relaxation process
and is described by the first-order differential equation due to the
Debye model of nonlinearity \cite{Akhm}
\begin{eqnarray}
t_{nl} \frac{d \delta n}{d t}+ \delta n=n_2 I, \label{relax}
\end{eqnarray}
where $n_2$ is Kerr nonlinear coefficient, $t_{nl}$ is the
characteristic relaxation time. We consider fast relaxing media
(electronic Kerr mechanism) with relaxation times as small as few
femtoseconds. Representing field strength as $E=A (t,z) \exp{[i(
\omega t-kz)]}$, where $\omega$ is a carrier frequency, $k=\omega
/c$ is the wavenumber, and introducing new, dimensionless arguments
$\tau=\omega t$ and $\xi=kz$, we come to the wave equation for the
pulse amplitude $A (t,z)$,
\begin{eqnarray}
\frac{\partial^2 A}{\partial \xi^2}-\frac{\partial^2 n^2 A}{\partial
\tau^2}-2 i \frac{\partial A}{\partial \xi}-2 i \frac{\partial n^2
A}{\partial \tau}+(n^2-1) A = 0. \label{Maxdl}
\end{eqnarray}
Usually, second-order derivatives are neglected at this point
resulting in slowly varying envelope approximation. However, in the
case of abrupt changes of refractive index (photonic crystal) it may
fail, so that we should solve the full Eq. (\ref{Maxdl}). In
addition, this equation allows to describe the behavior of electric
field in the structure without division into forward and backward
waves. Equation (\ref{Maxdl}) and the computational scheme
considered below are similar to those of Refs. \cite{Cren96, Novit}
where they were implemented to consider light propagation in a dense
resonant medium and a photonic crystal containing it.

Equation (\ref{Maxdl}) can be solved numerically by using the
finite-difference time-domain (FDTD) approach. The computational
scheme is based on calculation of amplitude value at every mesh
point $(l\Delta\tau, j\Delta\xi)$ as
\begin{eqnarray}
A_j^{l+1}=[-a_1 A_j^{l-1}+b_1 A_{j+1}^l+b_2 A_{j-1}^l+f A_j^l]/a_2.
\label{scheme}
\end{eqnarray}
Here the auxiliary values are
\begin{eqnarray}
a_1=(n_j^{l-1})^2 (1-i\Delta\tau), \qquad a_2=(n_j^{l+1})^2
(1+i\Delta\tau), \nonumber \\
b_1=\left(\frac{\Delta\tau}{\Delta\xi}\right)^2 (1-i\Delta\xi),
\qquad
b_2=\left(\frac{\Delta\tau}{\Delta\xi}\right)^2 (1+i\Delta\xi), \nonumber \\
f=2(n_j^l)^2-2\left(\frac{\Delta\tau}{\Delta\xi}\right)^2+\Delta\tau^2[(n_j^l)^2-1].
\nonumber
\end{eqnarray}

The values of refraction index at the mesh points can be obtained in
terms of finite-difference representation of Eqs.
(\ref{refr})-(\ref{relax}),
\begin{eqnarray}
n_j^{l+1}=n_{0j}+ \delta n_j^{l+1}, \nonumber \\
\delta n_j^{l+1}=\frac{\tau_{nl}}{\tau_{nl}+\Delta \tau} \left[
\delta n_j^l + \frac{\Delta \tau}{\tau_{nl}} n_{2j} |A_j^l|^2
\right],
\end{eqnarray}
where $\tau_{nl}=\omega t_{nl}$, $n_{0j}$ and $n_{2j}$ are the
values of the background refraction index and nonlinearity
coefficient at $\xi_j=j\Delta\xi$.

To correctly set the boundary conditions we use the total field /
scattered field (TF/SF) method and the perfectly matched layer (PML)
method which allows to apply the so-called absorbing boundary
conditions at the edges of the calculation region \cite{Anantha}.

\section{\label{SecTrap}Pulse trapping effect}

\begin{figure}[t!]
\includegraphics[scale=0.5, clip=]{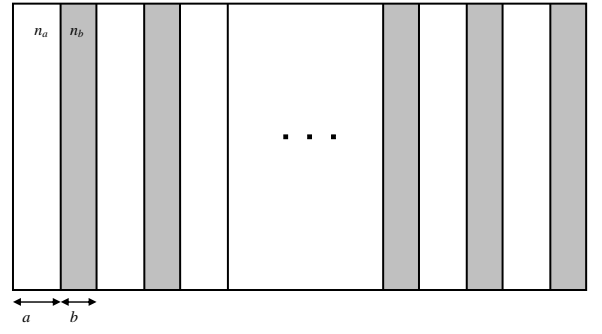}
\caption{\label{fig1} Scheme of a photonic crystal considered.
Parameters of it: refractive indices $n_a=2$, $n_b=1.5$; thicknesses
$a=0.4$, $b=0.24$ $\mu$m; number of layers $N=200$.}
\end{figure}

\begin{figure}[t!]
\includegraphics[scale=0.85, clip=]{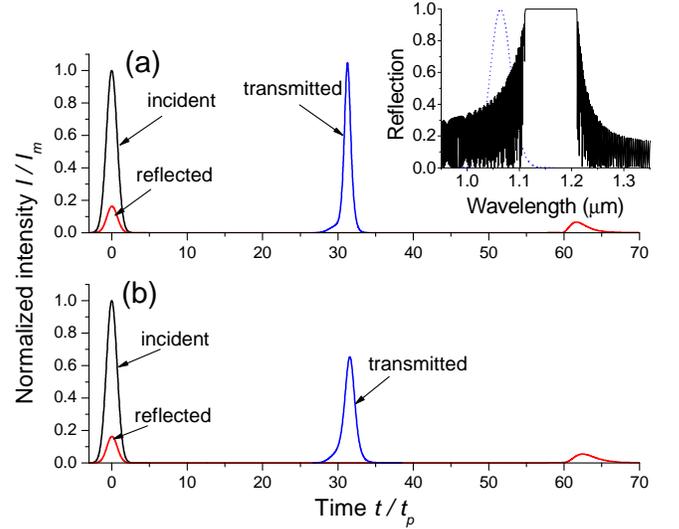}
\caption{\label{fig2} (Color online) Interaction of ultrashort pulse
with the photonic crystal with (a) relaxation-free ($t_{nl}=0$) and
(b) relaxing ($t_{nl}=6$ fs) cubic nonlinearity. Pulse peak
intensity $I_m=I_0$, where $I_0$ is such that $n_2 I_0=0.005$. Other
parameters are the same as in the caption of Fig \ref{fig1} and in
the text of the paper. The inset demonstrates the spectra of the
photonic crystal (solid line) and of the pulse (dotted line).}
\end{figure}

\begin{figure}[t!]
\includegraphics[scale=0.8, clip=]{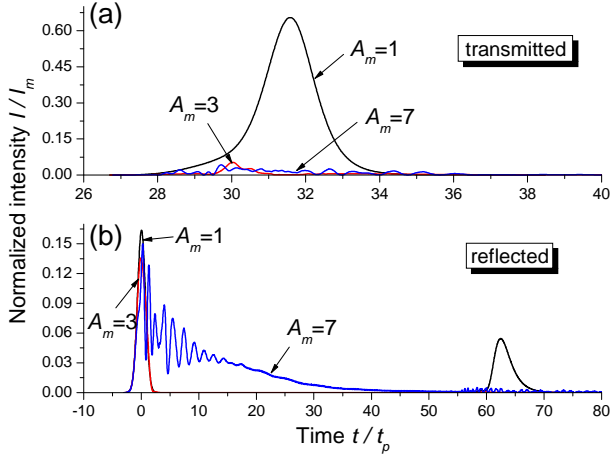}
\caption{\label{fig3} (Color online) (a) Transmitted and (b)
reflected intensity after interaction with nonlinear photonic
crystal at different incident pulse peak intensities ($A_m=1$, $3$,
$7$ in units of $A_0$). Relaxation time of nonlinearity $t_{nl}=6$
fs. Other parameters are the same as in the caption of Fig
\ref{fig2}.}
\end{figure}

\begin{figure}[t!]
\includegraphics[scale=0.9, clip=]{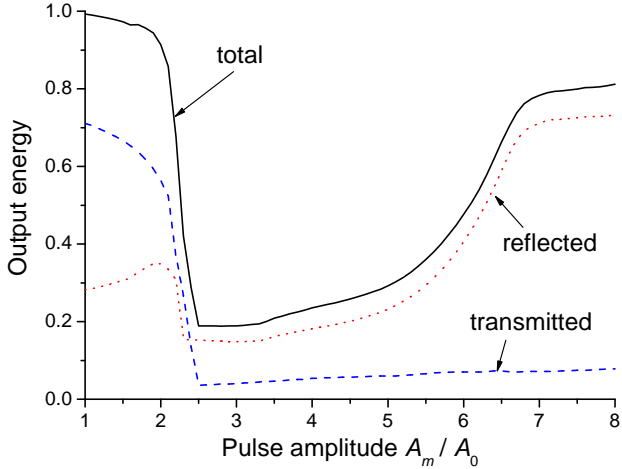}
\caption{\label{fig4} (Color online) Dependence of transmitted,
reflected and overall output energy (as fraction of input energy) on
incident pulse peak amplitude $A_m$. Relaxation time of nonlinearity
$t_{nl}=6$ fs. Energy was integrated over the time interval of $200
t_p$.}
\end{figure}

\begin{figure}[t!]
\includegraphics[scale=0.9, clip=]{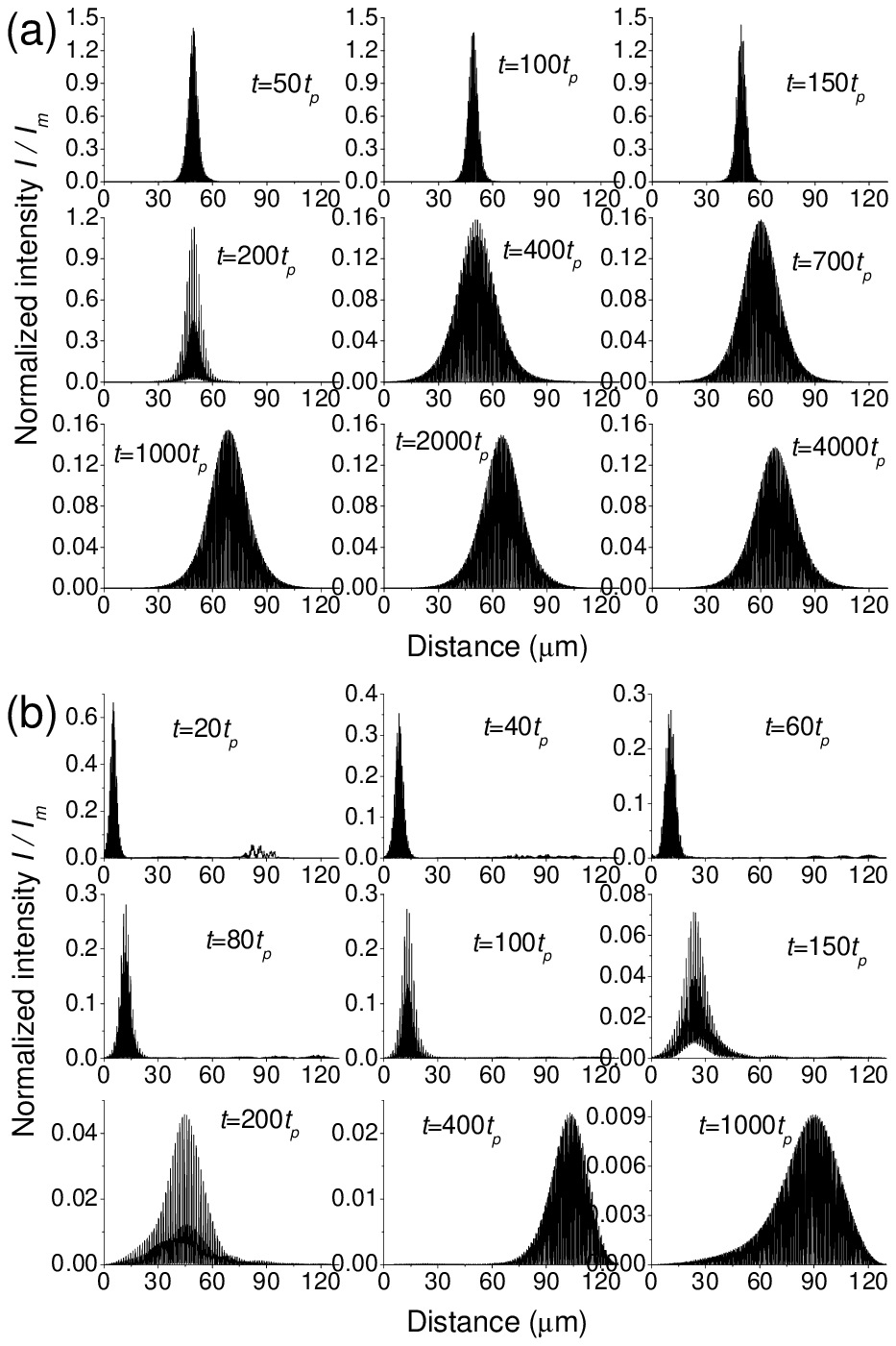}
\caption{\label{fig5} Distribution of light intensity inside the
photonic crystal at different time points. Pulse peak amplitude (a)
$A_m=3 A_0$, (b) $A_m=7 A_0$. Relaxation time of nonlinearity
$t_{nl}=6$ fs.}
\end{figure}

\begin{figure}[t!]
\includegraphics[scale=0.9, clip=]{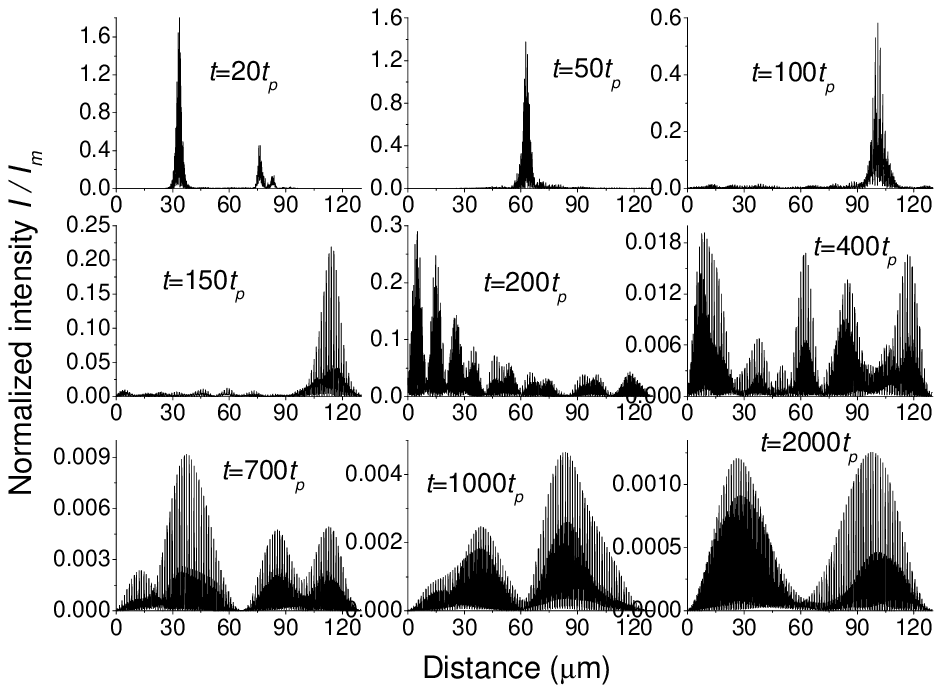}
\caption{\label{fig6} Distribution of light intensity inside the
photonic crystal at different time points. Pulse peak amplitude
$A_m=5 A_0$. Relaxation time of nonlinearity $t_{nl}=0$.}
\end{figure}

Let us consider propagation of ultrashort (femtosecond) light pulse
in one-dimensional photonic crystal shown in Fig. \ref{fig1}.
Spatial periodic modulation of the background refractive index $n_0
(z)$ defines the structure of it, so that it can be treated as a set
of alternate layers. The parameters used in calculations are as
follows: refractive indices of the layers $n_a=2$, $n_b=1.5$; their
thicknesses $a=0.4$, $b=0.24$ $\mu$m; number of layers $N=200$.
Nonlinear coefficient of the material is defined as $n_2 I_0=0.005$,
i.e. the pulse amplitude is normalized by the value
$A_0=\sqrt{I_0}$. This value of cubic coefficient provides
refraction index change of thousandth and hundredth of unity and was
used in calculations of Ref. \cite{Vlasov}. Note that nonlinear
coefficient is positive (focusing nonlinearity), so that refractive
index increases with the intensity. The incident pulse is assumed to
have Gauss envelope $A=A_m \exp{(-t^2/2t_p^2)}$. Here $t_p$ is a
pulse duration which further takes on the value of $30$ fs, while
the carrier frequency lies on the wavelength $\lambda=1.064$ $\mu$m.

Figure \ref{fig2} shows the results of calculations of pulse
interaction with nonlinear photonic structure with and without
relaxation. One can easily see that the effect of pulse compression
(obtained at $t_{nl}=0$) is completely absent when $t_{nl}=6$ fs.
This result is in strict accordance with the conclusions of Ref.
\cite{Vlasov}.

Vanishing of compression effect in Fig. \ref{fig2} was obtained for
pulse peak intensity $I_m=|A_m|^2=I_0$, where $I_0=|A_0|^2$ is the
value of intensity corresponding to $n_2 I_0=0.005$. If we take
greater $A_m$, the transmitted pulse continues to decrease. As it is
seen in Fig. \ref{fig3}(a), only small part of incident light can
pass through the nonlinear photonic crystal at $A_m=3 A_0$. This
situation is observed at $A_m=7 A_0$ as well, though the reflected
pulse gets larger in this last case [Fig. \ref{fig3}(b)]. When we
integrate intensity of reflected and transmitted light over a
certain time interval ($200 t_p$ in our calculations, i.e. large
enough in comparison with time required for pulse to pass through
the photonic crystal which is about $30 t_p$, see Fig. \ref{fig2}),
we obtain the characteristic energy curves (Fig. \ref{fig4}) that
describe the change in behavior of pulses as their peak intensity is
increasing. It is seen that, when $A_m$ gets larger than $2 A_0$,
reflected, transmitted and overall (summarized) energies demonstrate
dramatic decrease. After reaching the minimum (output energy is
about $20$ percent of the input one), the curves begin to rise
slowly. This slow growing of transmitted energy continues, while,
for reflected and total ones, rise becomes more steep (especially,
when $A_m>5 A_0$) and, finally, the plateau is observed for
$A_m\geq7 A_0$. Thus, the input energy almost entirely transforms to
reflected light at high intensities of incident pulse.

It is obvious that the energy of pulses with amplitudes between $A_m
\approx 2 A_0$ and $A_m \approx 6 A_0$ is confined inside the
photonic crystal. Figure \ref{fig5}(a) demonstrates light intensity
distributions along the length of the structure at different
instants of time for the pulse peak amplitude $A_m=3 A_0$. At first
($t=50 t_p$), the most part of pulse energy is localized in a narrow
region of the nonlinear photonic crystal, near the position $L=50$
$\mu$m (the total length of the structure is about $130$ $\mu$m). As
time goes by, energy tends to redistribute more uniformly.
Nevertheless, there is still pronounced maximum of intensity
distribution, moreover it is shifted towards larger positions,
namely $60-70$ $\mu$m at $t>700 t_p$. At large time points this
distribution stays almost invariant, or stationary. Its maximum only
slightly decreases, which seems to be connected rather with further
redistribution than with output radiation. Anyway, even at $t=4000
t_p$ approximately $80$ percent of pulse energy is still confined
inside the photonic structure, just as at $t=200 t_p$ (see Fig.
\ref{fig4}). This time is more than 100 times higher than the
interval needed for pulse to pass through the system. Recalling that
$t_p=30 fs$, this delay time in absolute units is greater than $100$
ps. Therefore we can say about \textit{pulse trapping} in this case.
Only for $t>4000 t_p$ the system starts to slowly emit light so that
the sharp distribution shown in Fig. \ref{fig5}(a) becomes violated.
More detailed calculations show that even at $t=10000 t_p$ about
$50$ percent of the initial energy is still inside the photonic
crystal, though it is distributed much more uniformly.

Now let us consider the pulse with $A_m=7 A_0$. The corresponding
spatial distributions are shown in Fig. \ref{fig5}(b). It turned out
that in this case the pulse is localized near the very beginning of
the structure. In this position it rapidly loses energy which is
mainly radiated through the front (input) end of the system. So this
radiation gives significant contribution to reflection [see Fig.
\ref{fig3}(b)]. After light intensity becomes lower than a certain
threshold, the pulse starts moving and widening. It moves quite
slowly, so that some part (about $20$ percent) of the input energy
is confined inside the photonic crystal for a long time, but the
peak intensity of the distribution is very low if we compare with
the case of $A_m=3 A_0$.

It is interesting to compare pulse behavior considered with pulse
propagation in photonic crystal with relaxation-free nonlinearity,
i.e. at $t_{nl}=0$. Intensity distributions inside the system for
this case are demonstrated in Fig. \ref{fig6}. It is seen that there
is no any long-term energy localization in non-relaxing photonic
structure. Light exhibits only chaotic "wander" inside it and
simultaneous attenuation due to emitting through input and output
ends. Finally, almost all energy of the pulse is already radiated by
the instant of time $t=700 t_p$. Therefore we can say that
relaxation of nonlinearity is a necessary condition to obtain the
effect of pulse trapping inside a photonic crystal.

\section{\label{SecMech}Physical mechanism of pulse trapping}

\begin{figure}[t!]
\includegraphics[scale=0.85, clip=]{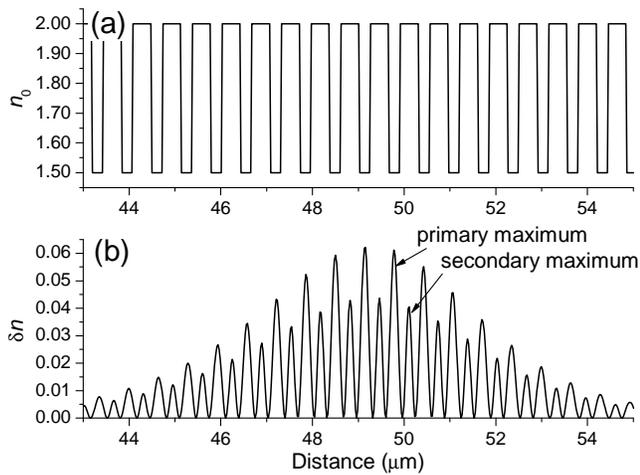}
\caption{\label{fig7} Spatial variation of (a) linear part of
refractive index $n_0(z)$, (b) nonlinear contribution $\delta n$.
Pulse peak amplitude $A_m=3 A_0$, time point $t=50 t_p$. Relaxation
time of nonlinearity $t_{nl}=6$ fs.}
\end{figure}

\begin{figure}[t!]
\includegraphics[scale=0.85, clip=]{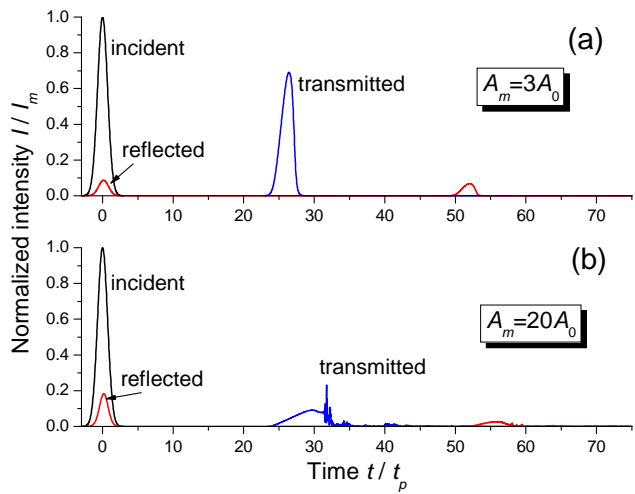}
\caption{\label{fig8} (Color online) Transmitted and reflected
pulses after interaction with a uniform cubic medium. Pulse peak
amplitude (a) $A_m=3 A_0$, (b) $A_m=20 A_0$. Linear part of
refractive index $n_0=1.8125$. Relaxation time of nonlinearity
$t_{nl}=6$ fs.}
\end{figure}

\begin{figure}[t!]
\includegraphics[scale=0.8, clip=]{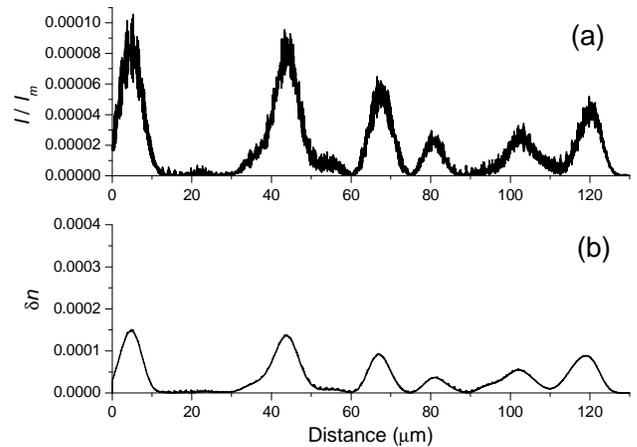}
\caption{\label{fig9} Spatial variation of (a) intensity, (b)
nonlinear variation $\delta n$ in the case of uniform cubic medium
with linear part of refractive index $n_0=1.8125$. Pulse peak
amplitude $A_m=20 A_0$, time point $t=1000 t_p$. Relaxation time of
nonlinearity $t_{nl}=6$ fs.}
\end{figure}

What is the physical reason, or mechanism, of this phenomenon? As
pulse propagates inside the photonic crystal possessing cubic
nonlinearity, refractive index of the structure changes dynamically
according to light intensity. If nonlinearity is relaxation-free,
these changes are instantaneous and depend entirely on field
distribution at current instant of time. In the case of relaxing
nonlinearity, nonlinear variation of refractive index ($\delta n$ in
our notation) can form a certain stable structure due to retardation
in its change. Appearance of this nonlinear dynamical "cavity"
results in pulse trapping: light tends to change the distribution of
$\delta n$ and leave the cavity, but inertia of nonlinearity
stabilizes it so that intensity is transformed to provide steady
spatial distribution $\delta n (z)$. The example of this is shown in
Fig. \ref{fig7}(b). One can see that maximal values of nonlinear
variation of refractive index $\delta n$ (so-called primary maxima)
are achieved in the layers with low linear refractive index
$n_0=1.5$ [see Fig. \ref{fig7}(a)]. This is in accordance with the
effect of light concentration in low refractive index regions of
photonic crystal for radiation tuned to high-frequency side of
reflection spectrum \cite{Bert} as it is in the case considered (see
the inset in Fig. \ref{fig2}). On the other hand, high refractive
index layers of the structure contain minor (secondary) maxima of
$\delta n$.

Obviously, low-intensity initial pulses are insufficient to produce
high enough nonlinear contribution $\delta n$, so that pulse is
mainly transmitted and reflected during short time after pulse
incidence. For higher intensities [e.g. $A_m=3 A_0$ as in Fig.
\ref{fig5}(a)] pulse forms nonlinear cavity inside the photonic
crystal which localizes the most part of pulse energy. Though this
cavity slowly tends to uniformity of $\delta n$ and loses its
energy, it allows to trap pulse light for relatively large time
interval. This behavior is obtained in wide range of pulse
amplitudes, approximately from $2.5 A_0$ to $5 A_0$. As pulse
intensity increases, cavity formation position moves towards the
front end of the photonic structure. Finally, for $A_m \approx 7A_0$
the nonlinear cavity appears so close to the entrance of the system
[Fig. \ref{fig5}(b)] that it rapidly emits almost all its energy in
the form of reflected light. That is why only intermediate
intensities of pulse (not too low and not too high) are suitable to
obtain the effect of trapping pulse inside the structure considered.

Finally, we should explain the role of photonic crystal in this
process. Is it necessary to use photonic-band-gap structure or,
maybe, the effect of pulse trapping can be observed in a uniform
cubic medium? To clarify this question we performed calculations of
pulse interaction with the medium possessing relaxing nonlinearity
and mean refractive index $n_0(z)=(a n_a+b n_b)/(a+b)=1.8125$. One
can see [Fig. \ref{fig8}(a)] that in this case there is no any light
localization for pulse amplitude $A_m=3 A_0$. Only much more
intensive pulses start to lose considerable part of energy inside
the medium. For example, about $40$ percent of pulse energy is found
to be trapped for $A_m=20 A_0$ [Fig. \ref{fig8}(b)]. This part of
energy stays inside the uniform medium for a long time and is
connected with corresponding nonlinear variation of refractive index
(see Fig. \ref{fig9} for the instant of time $t=1000 t_p$). However,
the distribution of light intensity in the uniform medium seems to
be stochastic and does not resemble pulse envelope as in the case of
nonlinear photonic crystal [compare the distributions in Figs.
\ref{fig9}(a) and \ref{fig5}(a)]. Therefore we cannot call light
localization in the uniform medium with relaxing nonlinearity by the
pulse trapping in the full sense of this term. Moreover, in photonic
crystals we need pulse intensities which are less by an order of
magnitude than in the case of the uniform medium.

\begin{figure}[t!]
\includegraphics[scale=0.85, clip=]{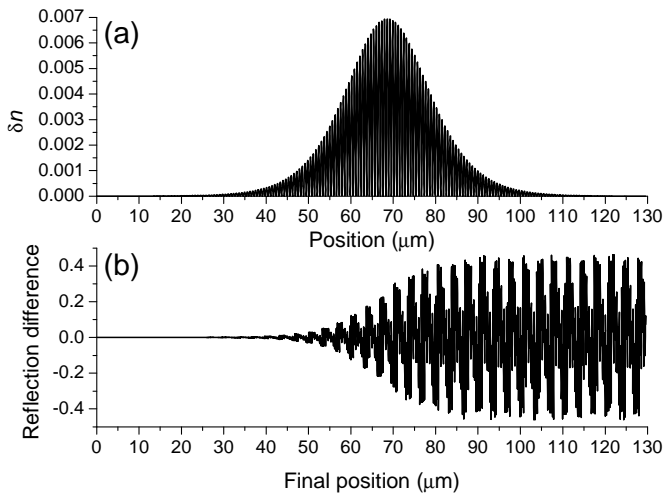}
\caption{\label{fig9_1} (a) Distribution of nonlinear refractive
index variation $\delta n$ corresponding to intensity distribution
of Fig. \ref{fig5}(a) at $t=1000 t_p$. (b) Difference between
reflection coefficients of the partial structures including the
layers from the input to a certain final position in the cases of
$\delta n$ given by picture (a) and $\delta n=0$. Calculations were
carried out for the central wavelength $\lambda=1.064$ $\mu$m.}
\end{figure}

Perhaps, the nonlinear cavity formation is connected with local
change of reflective properties of photonic crystal. This results in
dynamical shift of the band spectrum of the structure. To examine
this situation we consider the refractive properties (for the
central wavelength $\lambda=1.064$ $\mu$m) of the structure with
refractive index variations shown in Fig. \ref{fig9_1}(a) [it
corresponds to intensity distribution of Fig. \ref{fig5}(a) at
$t=1000 t_p$]. We calculate reflectivity of the partial structures,
which include the layers from the input to a certain final position
(inside the whole photonic crystal), and compare it with the case of
linear structure. Difference between reflectivities in this two
cases as a function of final position is demonstrated in Fig.
\ref{fig9_1}(b). It is seen that the structure with modified
refractive index modulation provides locally large reflectivity
deviations from the linear case. This deviations mainly appear at
large positions inside the crystal where $\delta n$ is high enough,
so that they can prevent light propagation in forward direction.
Similar situation is observed if we consider backward propagation.
Hence, it turns out that light is trapped in the central region of
the structure. Finally, it is worth noting that this picture of
refractive index variations and reflectivity deviations is
permanently changing, though it is stabilized by the relaxing
properties of nonlinearity.

\section{\label{SecDep}On optimal conditions of pulse trapping}

\begin{figure}[t!]
\includegraphics[scale=0.85, clip=]{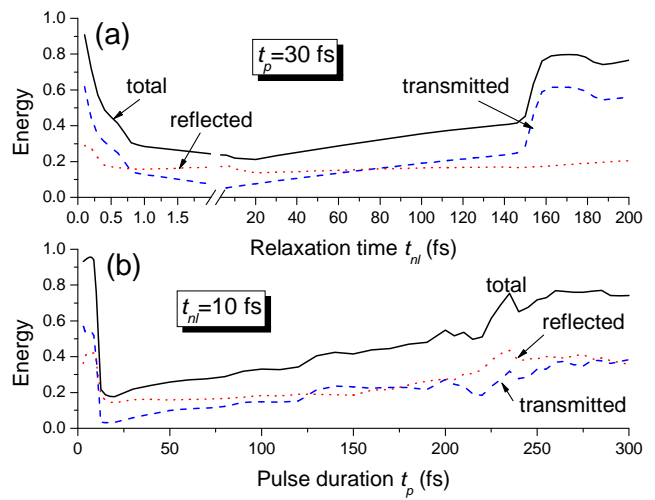}
\caption{\label{fig10} (Color online) Dependence of transmitted,
reflected and overall output energy (as fraction of input energy) on
(a) relaxation time at $t_p=30$ fs, (b) pulse duration at
$t_{nl}=10$ fs. Incident pulse peak amplitude $A_m=4 A_0$. Energy
was integrated over the time interval of $200 t_{p0}$, where
$t_{p0}=30$ fs.}
\end{figure}

\begin{figure}[t!]
\includegraphics[scale=0.95, clip=]{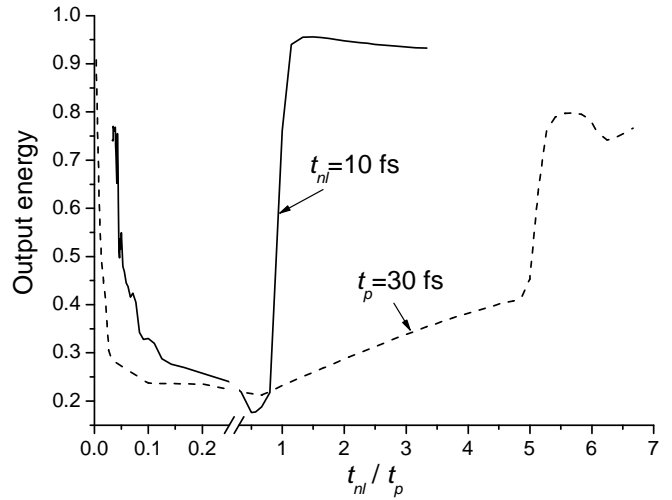}
\caption{\label{fig11} Dependence of overall output energy on the
ratio $t_{nl}/t_p$ plotted according to the data of Fig.
\ref{fig10}.}
\end{figure}

\begin{figure}[t!]
\includegraphics[scale=0.9, clip=]{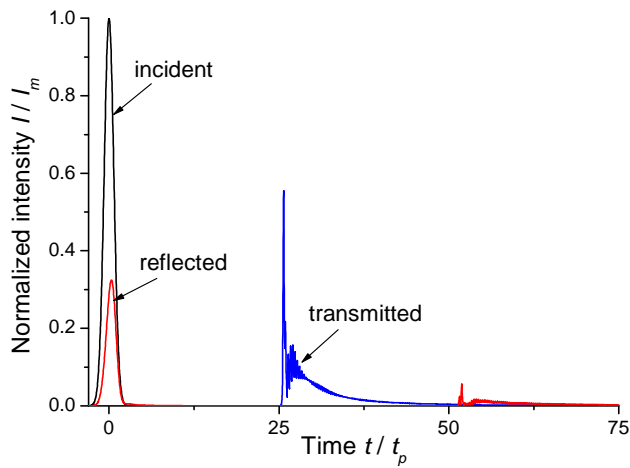}
\caption{\label{fig12} (Color online) Transmitted and reflected
pulses after interaction with a photonic crystal with negative
nonlinearity coefficient $n_2 I_0=-0.005$. Pulse peak amplitude
$A_m=3 A_0$. Relaxation time of nonlinearity $t_{nl}=6$ fs. Other
parameters are the same as in the caption of Fig \ref{fig1}.}
\end{figure}

In previous sections we considered the effect of pulse trapping in
photonic crystal with relaxing cubic nonlinearity only for a single
set of time parameters: $t_{nl}=6$ fs and $t_p=30$ fs. Fig.
\ref{fig10}(a) shows the dependence of output energies (reflected,
transmitted and total) on relaxation time at fixed value of pulse
duration $t_p=30$ fs. The behavior of these dependencies is similar
to that of the curves in Fig. \ref{fig4}: abrupt drop in the range
of small $t_{nl}$ (less than $1$ fs), smooth increasing and,
finally, steep rise of the curves for reflected and total energies.
The full range of relaxation times where the pulse trapping can be
observed is rather wide: from a fraction of femtosecond (relaxation
is so fast that it does not influence the pulse) to about $150$ fs
that is much greater than $t_p$ (medium reacts so slowly that the
nonlinear cavity forms near the very entrance of the system). The
optimal value is $t_{nl} \approx 10$ fs.

If we fix relaxation time $t_{nl}=10$ fs and vary pulse duration,
the behavior of output energies is quite different [Fig.
\ref{fig10}(b)]. As in the previous case, it demonstrates abrupt
decrease for small times, namely, pulse durations $t_p\simeq10$ fs
which correspond to only a few optical cycles. Such very short pulse
is not able to create a stable nonlinear cavity. On the other hand,
if $t_p$ is increasing, there is only stepless growth of output
energies, reflected and transmitted ones being approximately of the
same magnitude. The reason is the same: nonlinear cavity is not
formed as long as a whole pulse cannot be placed inside the
structure. The optimal value of pulse duration is about $20$ fs.

Since the limits of pulse trapping region are due to different
reasons in the cases of fixed $t_p$ and $t_{nl}$, the width of this
region will be different, too. In Fig. \ref{fig11} we plotted the
curves of Fig. \ref{fig10} versus the ratio $t_{nl}/t_p$. This
figure shows explicitly that relaxation time can be varied in much
more wide range than pulse duration. This also implies that we can
use media with relatively slow relaxing nonlinearities to obtain the
effect of pulse trapping.

Another question is connected with the role of sign of nonlinearity.
So far we considered only positive nonlinearity coefficient such
that $n_2 I_0=0.005$. If we take $n_2 I_0=-0.005$ (defocusing
nonlinearity), there is no any symptoms of pulse trapping as one can
see in Fig. \ref{fig12} for the amplitude $A_m=3 A_0$. Though this
problem should be studied in detail, the preliminary conclusion is
that trapping can be observed only for $n_2>0$, at least for
comparatively low intensities.

\section{Conclusion}

To summarize, we have analyzed the possibility of trapping pulse in
photonic crystal with relaxing cubic nonlinearity. By using
numerical simulations, we showed that this process is due to the
balance between light spreading and inertia of nonlinearity which
results in steady nonlinear cavity formation and pulse trapping
within it. Photonic crystal is a necessary element for this cavity
to appear, due to the processes of dynamical local change of
reflection and transmission of the nonlinear structure, and, in
addition, leads to decreasing of pulse intensity required to observe
trapping. We discussed the reasons for pulse trapping disappearance
at high and low values of both pulse duration and relaxation time
resulting in existence of the range of optimal magnitudes of these
parameters.

\section*{Acknowledgements}

Author wants to thank Dr. Andrey Novitsky for fruitful discussions.
The work was partially supported by the State Complex Program for
Scientific Research "Photonics".


\begin{thebibliography}{17}
\bibitem{Yabl} E. Yablonovitch, \jmo {\bf41}, 173 (1994).
\bibitem{Bert} M. Bertolotti, J. Opt. A {\bf8}, S9 (2006).
\bibitem{Joan} J.D. Joannopoulos, R.D. Meade, and J.N. Winn, \textit{Photonic crystals} (Princeton University Press, Princeton, 1995).
\bibitem{Stal} K. Staliunas, C. Serrat, R. Herrero, C. Cojocaru, and J. Trull, \pre {\bf74}, 016605
(2006); Yu. Loiko, R. Herrero, and K. Staliunas, \josab {\bf24},
1639 (2007).
\bibitem{Chutinan} A. Chutinan, S. John, and O. Toader, \prl {\bf90}, 123901 (2003); A. Chutinan and S. John, \pre {\bf71}, 026605 (2005).
\bibitem{Vujic} D. Vujic and S. John, \pra {\bf72}, 013807 (2005).
\bibitem{Scalora} M. Scalora, M.J. Bloemer, A.S. Manka, J.P. Dowling, C.M. Bowden, R. Viswanathan, and J.W. Haus, \pra {\bf56}, 3166 (1997).
\bibitem{Ming00} S.F. Mingaleev, Yu.S. Kivshar, and R.A. Sammut, \pre {\bf62}, 5777 (2000).
\bibitem{Ming01} S.F. Mingaleev and Yu.S. Kivshar, \prl {\bf86}, 5474 (2001).
\bibitem{Ming02} S.F. Mingaleev and Yu.S. Kivshar, \josab {\bf19}, 2241 (2002).
\bibitem{Sukh} A.A. Sukhorukov and Yu.S. Kivshar, \pre {\bf65}, 036609 (2002).
\bibitem{Eggl} B.J. Eggleton, R.E. Slusher, C.M. de Sterke, P.A. Krug, and J.E. Sipe, \prl {\bf76}, 1627 (1996).
\bibitem{Zhelt} A.M. Zheltikov, N.I. Koroteev, S.A. Magnitskiy, and A.V. Tarasishin, Quantum Electron. {\bf 28}, 861 (1998).
\bibitem{Good} R.H. Goodman, R.E. Slusher, and M.I. Weinstein, \josab {\bf19}, 1635 (2002).
\bibitem{Mak1} W.C.K. Mak, B.A. Malomed, and P.L.Chu, \pre {\bf67}, 026608 (2003).
\bibitem{Mak2} W.C.K. Mak, B.A. Malomed, and P.L.Chu, \josab {\bf20}, 725 (2003).
\bibitem{Akhm} S.A. Akhmanov, V.A. Vysloukh, and A.S. Chirkin, \textit{Optics of Femtosecond Laser Pulses} (AIP Press, New York, 1992).
\bibitem{Vlasov} R.A. Vlasov and A.G. Smirnov, \pre {\bf61}, 5808 (2000).
\bibitem{Anantha} V. Anantha and A. Taflove, IEEE Trans. Antennas Propag. {\bf50}, 1337 (2002).
\bibitem{Cren96} M.E. Crenshaw, \pra {\bf54}, 3559 (1996).
\bibitem{Novit} D.V. Novitsky, \pra {\bf79}, 023828 (2009).

\end{thebibliography}
\end{document}